Article

**Brain waves are a repetition of a pause and an activity**

Chika Koyama

Department of Small Animal Clinical Sciences, School of Veterinary Medicine, Rakuno Gakuen University, Hokkaido, 069-8591, Japan

*Correspondence to Chika Koyama, Wako, Saitama, 351-0101, Japan
e-mail: chik_0711@icloud.com

**Declaration of Competing Interests**: The author has no conflicts to disclose.

**Abstract**

Brain waves still cannot reliably distinguish between awake and asleep states. Here, I present new original indices, voltage subthreshold wave "**τ**" and abovethreshold wave "**burst**", for advanced LFP/EEG readings. Assuming that **τ** is a microwave that fluctuates every sample such as the equipotential, the total number of **τ** (N**τ**) is inferred to be the maximum, and the amplitude of **burst** (Abst) is inferred to be the minimum. In fact, they invariably had a mean **τ** duration (M**τ**) of 2-3 sample intervals in any case. In addition, **τ** and **burst** exhibited self-similarity for sample frequency while occupying approximately 30% and 70% of LFP in the natural state, respectively. Its threshold and Abst were correlated with the vigilance state and decreased to 70% by doubling the sample frequency. The dose of sevoflurane, which inhibits and synchronizes neural activity, was linearly correlated with decreases in the threshold and N**τ**. Thus, **τ** could reflect the uncertainty of the membrane potential. I propose that **τ** and **burst** represent a pause and an activity such as the rhythm of the brain.

**Introduction**

One of the remaining issues in brain research will be the development of brain information reading technology with high precision (*1*). Electroencephalography (EEG) signals are generated mainly by the postsynaptic potential of pyramidal neurons in cortical layer 5 (*2-4*), and their activities play a core role in the mechanism for generating consciousness (*5, 6*).

EEG anesthesia depth monitors have been commercialized for more than a quarter century. However, their effectiveness in preventing intraoperative awareness has not yet been confirmed (*7*). Anesthetics are known to cause neuronal hyperpolarization by increasing inhibition and decreasing excitation (*8, 9*) and brain-wide synchrony in layer 5 pyramidal neurons of the cortico-thalamus loop (*10*). EEG changes from slow wave activity to burst-suppression (BS) activity with increasing anesthetic dose (*11*). Slow wave activity is observed in synchrony with long-lasting hyperpolarization, and their neurons are based on a bistable updown pattern alternating between hyperpolarization and depolarization (*12-14*). BS pattern is based on a unimodal state in which hyperpolarization is predominant (*11*).

To break through the individual variability that is a bottleneck in brain-wave research, I propose that brain waves include equipotential fluctuations in addition to oscillations. We can confirm that the brain wave contains flattish periods where there is not much variance in voltage between successive peaks. Magnifying the waveform shows that those flattish periods fluctuate every one or a few samples (called “microwave”), even when the sampling frequency is increased (ex. Fig. 1b-d). Since those fluctuations could be regarded as momentary equipotentials, I hypothesized that their amplitude could reflect ambient activity. Here, I classified brain waves into subthreshold waves (defined as "**τ**") and abovethreshold waves (defined as "**burst**") (Fig. 1a-d) and calculated the number of **τ** (N**τ**), mean duration of **τ** (M**τ**), and mean amplitude of **burst** (Abst) for every 1 min of data. This study was conducted using 125 Hz – 1000 Hz extracellular local field potential (LFP) at cortical layer 5 at the secondary motor cortex (M2) and primary somatosensory cortex (S1) in nonmedicated 6 mice and 256 Hz EEG from T4 and Fpz electrodes in 6 senior and 6 young adult dogs anesthetized with sevoflurane (SEV). Using the existing identification method, I first demonstrated that **τ** and **burst** indices can be used for identifying vigilance states. Next, examining the change pattern in the indices with increasing threshold from 1 μV to 100 μV, it was shown that the maximum N**τ** and the minimum Abst were always at M**τ** of approximately 2.5 sample intervals on both LFP and EEG. Finally, I examined the relationship between the threshold with a M**τ** of 2.5 sample intervals as the amplitude of microwaves and the vigilance state or anesthesia dose.

## Results

### Changes in Nτ and Abst with increasing LFP threshold in mice

The scatterplots of N**τ** and Abst for 20 h formed 2 distinct populations at most thresholds from 0.001 mV to 0.200 mV in all subjects (ex. Figs. 1e-h, 2a, 2d-f, 3a-b). In Nos. 1 and 2, those populations were almost perfectly consistent with the state identification of awake and NREM by the existing method (ex. Figs. 1e-h, 3a-b). There were no REM trials identified in Nos. 1, 2, and 5 by the existing method (Figs. 2a, 3a, Fig. S1g). Since the high-amplitude spike indicates awake (*15*), as shown in Fig. 1b-d, REM waveform was considered similar to NREM waveform in **τ** and **burst** analysis. Therefore, in the following, those two populations are assigned to awake and sleep states.

N**τ** increases and then decreases as the threshold increases (ex. Fig. 1e-f). This change in sleep state precedes that in awake state. At 1000 Hz LFP (Fig. 1f), N**τ** changes with smaller thresholds than that at 125 Hz (Fig. 1e) due to the fineness of the waveforms (Fig. 1b-d). At thresholds above 0.100 mV at 125 Hz and 250 Hz, N**τ** in the sleep state varies, as shown in Fig. 1e. This could be explained by the difference between the waveforms in NREM and REM; above a certain threshold, the low-amplitude continuous wave of REM would increase M**τ**. Then, its N**τ** becomes lower than that in NREM, especially at lower sampling frequencies, such as 125-250 Hz, because the slow wave of NREM is less subdivided at lower sampling frequencies (Fig. 1c, 1d). In contrast, Abst decreases and then increases as the threshold increases (ex. Fig. 1g-h). Abst of awake state tends to be greater than that of sleep, especially at higher sampling frequencies and lower thresholds where slow waves are more subdivided.

### Identification of 4 vigilance states using Nτ on 125 Hz LFP at M2

N**τ** of the 125 Hz LFP at M2 with a threshold of 0.150 mV (Nos. 1-2) or 0.100 mV (Nos. 3-6), where most N**τ**s of sleep state are approximately 200, were used for state identification. I set up two variations from the main sleep population (Figs. 2a, 3a): downward variation (called "smaller N**τ**") and upward variation (called "greater N**τ**"). Examples of **τ** and **burst** on LFPs are shown in Figs. 2b and 2c; Typical REM waveforms (ex. Fig. 2b_ left top) indicate that LFP is a collection of longer subthreshold waves, i.e., smaller N**τ** and larger M**τ** (Fig. 2a_bottom). Their total **τ** ratio, obtained by multiplying M**τ** by N**τ**, exceeds approximately 80% or more (Fig. S1h). The increased appearance of slow waves and spikes are accompanied by larger N**τ** and smaller M**τ**. LFP with greater N**τ** shows a mixture of NREM and awake waveforms (Fig. 2c, left top). Therefore, for the main sleep population of "NREM", the smaller N**τ** and the greater N**τ** can be used as indicators of "REM" and "light sleep", respectively.

Here, for the identification of 4 vigilance states (awake, light sleep, NREM, REM), hierarchical clustering was performed on N**τ** in two steps. The first step was that Ward's linkage of hierarchical clustering divided all trials into 2 classes: awake state (N**τ**: 588±91, (n=1791)) and sleep state, assigned in descending order of N**τ**. In the second step, centroid linkage (No. 1, Nos. 3-6) or average linkage (No. 2) of hierarchical clustering further divided the sleep state trials into 3 classes: light sleep state (N**τ**: 349±40, (n=183)), NREM state (N**τ**: 192±40, (n=2879)), and REM state (N**τ**: 75±33, (n=253)), assigned in descending order of N**τ.** These results are shown in Figs. 2a, 3b, and S1. The values of the indices in each state are summarized in Fig. S1h. As shown in Figs. 2g and 3c-f, N**τ**, M**τ**, and Abst are state-specific values at various thresholds, suggesting that any small waves are state specific.

The spectral edge frequencies of 95% (SEF95) of the awake at M2 in Nos. 1 and 2 are approximately twice as large, and those standard deviations are less than half of those in Nos. 3-6 (Fig. S2). The SEF95s are over 40% of the sampling frequency, and those standard deviations are less than 10% of the SEF. This means that awake LFP waves at M2 of Nos. 1 and 2 are consistently dominated by fast waves, while those at Nos. 3-6 often contain slow waves with fast waves. Therefore, the state identifications of Nos. 1 and 2, whose identifications are almost perfectly consistent with this classification using N**τ**, can be typically identified by frequency analysis. Thus, these results of SEF95 will also confirm the accuracy of **τ** and **burst**.

**Mτ of 2.2-2.9 sample intervals maximizes Nτ and minimizes Abst.**

The mean values of indices per vigilance state were calculated at every 1 µV threshold from 1 µV to 100 µV (Figs. 2g, 3c-f, Fig. S3). As the threshold increases, the mean M**τ** (called M**τ**) increases, the mean N**τ** (called N**τ**) increases and then decreases, and the mean Abst (called Abst) decreases and then increases. In all cases, the maximum N**τ** and the minimum Abst were at almost constant M**τ** values of 2.69±0.22 and 2.34±0.11 sample intervals, respectively (Fig. 4a, 4d). Thresholds that meet those small M**τ** can be estimated as representative values of the microwave amplitude. Then, its N**τ** and Abst are for the case where **τ** is a microwave.

Considering the uniform pattern of indicator changes with respect to threshold changes, for N**τ** to change from increasing to decreasing, separated **τ**s must be joined together, as shown in Fig. 4c. In contrast, for Abst to change from decreasing to increasing, the decrease in amplitude of **burst** fragmented by **τ** must be smaller than the increase in the amplitude of **burst** with increasing the threshold. The **burst** fragmentation can be established by **τ** of even one sample interval generated (Fig. 4c). In addition, **τ** in the middle of a large wave appears against the direction of the slope, and then its amplitude of **τ** is reduced by that amount of its inclination. This can explain why M**τ** at the peak of Abst was always smaller than that of N**τ**. An increasing **burst** requires small amplitude **burst**s to be converted to **τ**s as the threshold increases.

**When τ is a microwave, τ and burst exhibit self-similarity, region specificity, and state specificity.**

The thresholds for maximizing N**τ** (ThN**τ**max) and minimizing Abst (ThAmin) and their N**τ**, M**τ**, and Abst are specific to the sampling frequency (Fig. 4a and 4d). As the sampling frequency doubles (125 Hz -> 250 Hz -> 500 Hz -> 1000 Hz), ThN**τ**max, ThAmin and Abst are reduced to 60-80%, N**τ** is roughly doubled, and M**τ** is roughly halved. As shown in Fig.S4, **τ** with a duration of approximately 2.5 sample intervals generates every 8 to 13 samples on average.
ThN**τ**max and ThAmin were significantly greater at S1 than at M2 in all states and significantly greater in awake, light sleep, REM, and NREM states (Fig. 4a and 4d) in proportion to brain activity. Both ThN**τ**max and ThAmin were greater in REM than in NREM at S1 (n=48), whereas both ThNτmax and ThAmin were nearly equal between REM and NREM at M2. Since REM state is characterized by relaxation of skeletal muscles, the finding that the threshold increase in REM is not evident at M2 but is more pronounced at S1 strongly suggests a correlation between those thresholds and brain activity.
Absts at both ThN**τ**max and ThAmin were significantly greater at S1 than at M2 in all states and were significantly greater in the following order: awake, light sleep, NREM, and REM ($p<10^{-5}$) (Fig. 4a and 4c). The difference in Abst between REM and NREM LFP was obvious at 125-250 Hz, especially at 125 Hz, but not at 1000 Hz. This could be explained by the presence of microwaves on the slope of the slow wave that disappear at lower frequencies (Fig. 4f).
The ratio of total **τ**, which is N**τ** multiplied by M**τ**, is nearly constant even if the sampling frequency is changed (Fig. 4b and 4e). The mean ratios of total **τ** are 0.30±0.03 at M2 (n=24), 0.29±0.03 at S1 (n=24) at ThN**τ**max, 0.25±0.03 at M2 (n=24) and 0.25±0.03 (n=24) at S1 at ThAmin.

### Changes in Nτ and Abst with increasing threshold and SEV on EEG in dogs

With increasing threshold on EEG in dogs, N**τ** increases and then decreases, and Abst decreases and then increases (Fig. 5a, 5b and Figs. S5, S6), as does LFP in mice. As SEV increases, the maximum value of N**τ** decreases at non-BS level and then increases at BS level. The decreasing slope of N**τ** for the threshold also shows an SEV dose-specific gradient, which becomes more gradual at non-BS level and conversely becomes steeper at BS level. Conversely, as SEV increases, the minimum value of Abst increases at non-BS level and then decreases at BS level. Abst for the threshold increases constantly as SEV increases at non-BS level.
The correlation coefficients (r) between SEV of the suppression ratio (SR) <0.32 and N**τ** or Abst for the threshold are summarized in Fig. 5e and 5f. At thresholds of approximately 2 μV, N**τ** is strongly negatively correlated with the SEV. As thresholds increases above 4 μV, r increases and shifts to a strong positive correlation. At thresholds above 1 or 2 μV, Abst is strongly correlated with the SEV.
As SEV increases from 2.0% to 3.0%, the SEF95 is almost constant in the seniors of Nos.

1-6 and tends to decrease in the young adults of Nos. 7-12. At SEV of 5%, in the young adults except for Nos. 10 and 12, monotonous alpha activity was observed as the suppression wave of BS (ex. Fig. 5c_No. 11). SEF95 tends to be greater at pre-BS and BS levels due to spikes without accompanying slow waves (ex. Fig. 5c and 5 h).

**Effect of increasing SEV on τ of microwave and burst**

The mean Mτ of all SEVs with the maximum Nτ and minimum Abst was 10.3±2.1 (ms) and 8.9±1.2 (ms) in the seniors of Nos. 1-6 and 10.2±2.6 (ms) and 8.6±1.8 (ms) in the young adults of Nos. 7-12, respectively (Fig. 6b, 6d). The 2.7 sample intervals at 256 Hz corresponds to 10.547 ms, and the 2.3 sample intervals corresponds to 8.984 ms. Hence, the Mτs via EEG in dogs are almost the same as those via LFP in mice.

Next, I showed strong correlations between SEV and Abst, Nτ, or the threshold when SR is less than 0.15 and Mτ is approximately 2.3-2.7 sample intervals (tests 1-4; Fig. 6e-g). The r between SEV and Abst or Nτ did not significantly differ among the 4 tests (Fig. 6g): SEV vs Abst, r = 0.85 ± 0.04 and 0.93 ~ 0.94 ± 0.05; SEV vs Nτ, r = -0.82 ~ -0.86 ± 0.08 ~ 0.10 and -0.92 ~ -0.94 ± 0.03 ~ 0.05 (seniors and young adults). The r is greater for the young adults than for the seniors. In contrast, the r between SEV and the threshold is greater when M is aligned (tests 3 and 4) and in the seniors than in the young adults. In tests 3 and 4, r exceeds 0.9 for all seniors except for No. 3 and for young adults Nos. 10 and 12. In the young adults, except for Nos. 10 and 12, the thresholds are not lowered enough despite the higher SR at SEV of 5%. Their EEG waveform showed that alpha activities are on the suppression wave of BS (ex. Fig 5c_No.11). The threshold increases paradoxically, at SEVs of 2.5% and 3.5% in No. 3, at 3.5% in No. 9, and at 4.0% in Nos. 7 and 8.

**Discussion**

**τ** and **burst** indices show that brain wave data contain sufficient features to be constant across individuals and species for identification of the vigilance state and hypnotic level. I first demonstrated using the 125 Hz LFP of mice at a threshold where N**τ** in NREM is approximately 200, the state was judged as REM where N**τ** is 150 or less, light sleep where N**τ** is approximately 300, and wakefulness where N**τ** is 400 or more. The N**τ** values were uniform for all the mice even for various LFPs. Nevertheless, compared with NREM waves (M**τ** = 270 ± 86 ms), conventional state-specific waveforms can be also explained by a decrease in N**τ** due to the extension of **τ** (M**τ** = 880 ± 543 ms) on low-amplitude fast waves of REM and an increase in N**τ** due to the fragmentation of **τ** (M**τ** = 54 ± 21) by spikes on active waves during wakefulness (Fig. S1h).

Next, when **τ** is a microwave, **τ** and **burst** indices show robust features such as fractal structure as self-similarity of invariance to time scale changes (Fig. 4, Fig. S4): The maximum N**τ** and minimum Abst were found when M**τ** was 2.7 and 2.3 sample intervals, respectively, which is common to both mouse LFP and dog EEG data. In addition, N**τ** was approximately doubled with the frequency doubling. Therefore, at all frequencies, the ratio of the total **τ** (M**τ** multiply N**τ**) was constant at the maximum N**τ** and minimum Abst. This suggests that equipotential fluctuations have physical properties of neural activity, supporting the equipotential effect of **τ** on the fractal landscape. Moreover, its threshold and Abst correlated with the vigilance state and decreased to approximately 70% by doubling the sampling frequency, supporting the active potential of **burst** on fractal landscape. Therefore, I propose that **τ** and **burst** would be confirmed as units of a pause and an activity, similar to brain wave rhythms.

Letting **τ** be a pause, decreasing N**τ** with increasing sevoflurane dose (Fig. 6 e-g) indicates that sevoflurane causes fewer pauses and longer periods of activity. General anesthetics are known to synchronize the neural activity of layer 5 in the cortico-thalamus loop (*10*) while suppressing action potential firing especially at high frequencies (*15-17*). This synchronized activity may be consistent with the less pauses and longer activity. An increase in Abst inversely proportional to N**τ** (Fig. 6i) can be interpreted as the result of prolonging the activity with fewer pauses rather than the increasing slow waves and spikes: the maximum percentage of slow waves in young adults was assumed to be at 2.5%-3.0% of SEV with the minimum SEF95 (Fig. 5d), whereas the maximum Abst was at 3.5%-4.0% of SEV (Fig. 6f). The effect of spikes at the pre-BS level should be also not noticeable; the Abst had an average amplitude of 600 ~ 1000 **burst**s, while the number of spikes was only 100 at most (ex. Fig. 6f).

In two-third of young adult dogs, the paradoxical increases in threshold and N**τ** occurred with the advent of BS. Such a paradoxical trend in the hypnotic index during deep anesthesia has also been reported with commercially available EEG hypnosis monitoring (*18-20*). The onset of BS should be unstable as a transient phase from slow wave in bistable state alternating between hyperpolarization and depolarization (*12-14*) to long suppression in predominantly hyperpolarized unimodal state (*11*). Indeed, it has been reported that cortical hyperexcitability caused by suppressed neuronal inhibition is involved at the transition (*14, 21*). In contrast, the decreases in the threshold and N**τ** with increasing SEV (Fig. 6e-g) could also indicate that SEV causes a decrease in the amplitude and probability of equipotential **τ**. Therefore, those paradoxical increases indicate that instability at the onset of BS affects the amplitude and probability of equipotential **τ.** Thus, the equipotential **τ** may be correlated with the uncertainty of the membrane potential. This could be a reason for the changes in the threshold and N**τ** in this study.

Previous reports have shown that anesthetics induce robust linear dose-dependent attenuation of cortical power in the 76-200 Hz range using LFP amplification at 0.1-475 Hz (*22, 23*). Since the 76-200 Hz range corresponds to the 2-6 sample intervals, the linear attenuation of the power is roughly the same as the decrease in the threshold and N**τ** with increasing SEV.

In all the mice, REM sleep-like waveforms were not observed during bright hours (Figs. 2a_middle, 3b). This may be explained by the fact that blue light inhibits REM-like waveform formation (*24*). As shown in No.3 of Figure 3a and 3b, remarkable collective deviations of indices are observed in trials of less than 150. It seems that the magnitude of the voltage acquisition changed with the passage of time, which may indicate that there were fluctuations in the resistance values due to biological changes around the electrode. This requires further investigations.

## Materials and Methods

### 1. Experimental animals and ethics

All mouse experiments were performed in accordance with institutional guidelines and were approved by the Animal Experiment Committee of the RIKEN Brain Science Institute (H24-2-231(6)). Six Thy1-ChR2 (Jackson Laboratory, Bar Harbor. ME, USA No. 1-No. 6), maintained on a C57BL/6JJmsSlc background, and CA1-specific Cre mouse line (CaMKIIα-Cre;CW2) mice (*25*) were used to measure LFP. In all experiments, the mice were housed under a 12 h light:12 h dark (light on: 8 AM/light off: 8 pm) cycle in individually ventilated cages with 1-2 animals per cage.

All dog experiments were performed in accordance with institutional guidelines and were approved by the Animal Care and Use Committee of Rakuno Gakuen University (No. VH14B7). Six healthy senior beagles (10.1 ± 1.5 years old; 12.6 ± 1.4 kg; No. 1-6) and 6 healthy young adult beagles (2.5 ± 1.5 [mean ± SD] years old; 9.9 ± 0.9 kg; No. 7-12), 3 males and 3 females each, were subjected to EEG under anesthesia with 2.0% to 5.0% sevoflurane (SEV) and muscle relaxants. The 2.0% concentration was determined based on the reported value of 1.3 ± 0.3% for the minimum alveolar concentration preventing voluntary response in 50% of dogs (MAC-awake) of SEVs (*26*).

### 2. Data acquisition (LFPs in mice)

Continuous LFP recordings were performed using 75-µm platinum electrodes from M2 and S1 in the right hemisphere. Continuous EMG recordings were performed through a slip ring. To target cortical layer 5, M2 and S1 electrodes were inserted at a depth of 670 µm. These parameters were recorded 24 hours a day from 0 AM. Electrical signals were filtered at 0.1 Hz-5 kHz with an amplifier and digitized at 10 kHz. LFPs at M2 and EMGs were used for computer-based online sleep scoring via an existing method (*27*) (*28*). LFP data at M2 and S1 were used to examine **τ** and **burst**. A total of 1200 LFP data sets of 60 s were extracted for analysis per mouse from the data of 20 hours, excluding the 2 hours each after the start and before the end of recording. The data that were determined to be in the same state for all 60 s were used for **τ** and **burst** analysis and frequency analysis. I used LFP data for which protocols were previously reported (*29*).

### 3. Data acquisition (EEGs in dogs)

In order at SEVs of 2.0%, 2.5%, 3.0%, 3.5%, 4.0% and 5.0%, following 20 min of equilibration, the raw EEG data (256 Hz sampled) measured by an A-2000XP BIS monitor (Covidien-Medtronic, MN, USA) were recorded for 5 min using a “Bispectral Analyzer (BSA) for A2000” software (*30*). Avoiding the data including checking time of the monitor, three 64 s EEG data packets for analysis were separated from 5 min of raw data. To preprocess the data, they were divided into nonoverlapping 2 s periods. Then, the power spectral density was estimated by fast Fourier transformation after applying Welch's window function. Subsequently, discrete Fourier transformation was performed at 50 ± 1 Hz and 100 ± 1 Hz to remove the noise from the alternating current. I used previously reported EEG data (*31*).

#### 4. τ and burst analysis

The peaks of LFP and EEG waves were detected through their first-order derivatives. The threshold value was set every 0.001 mV from 0.001 mV to 0. 100 mV, and 0.125, 0.150, 0.200 mV for 60 s LFP, and every 0.05 μV from 0.05 μV to 10 μV for 64 s EEG. A subthreshold period where one-half of the adjacent peak potential difference does not exceed the threshold is defined as **τ**, and an above-threshold period where it exceeds the threshold is defined as **burst** (Fig. 1a). Based on **τ** and **burst** components on a 60 s LFP or 64 s EEG, 3 new parameters were developed as follows:

**τ: subthreshold period**

**Nτ = total number of τ**

**Mτ (s) =** $\mathbf{\Sigma\ \tau\ /\ N\tau}$

**burst: above-threshold period**

**Abst (mV or μV) =** $\mathbf{\Sigma\ amp\ \ /\ Nbst}$

$\Sigma\ \boldsymbol{\tau}$ denotes the sum of all **τ** events. Nbst is the total number of **burst** events and is equal to N**τ** or N**τ** ± 1. $\Sigma\ \mathrm{amp}$ denotes the sum of the difference between the maximum and minimum voltages in a **burst**. M**τ** (ms) is the mean **τ** duration. Abst (mV) is the mean **burst** amplitude. The data were analyzed by C program with custom-made scripts.

#### 5. SEF95

One-second moving averages were subtracted from the data, which were divided into nonoverlapping 2-s periods, and Welch's window function was applied. Subsequently, discrete Fourier transform was performed at 50 ± 1 Hz and 100 ± 1 Hz to remove noise caused by the alternating current power source. The power spectrum density was estimated by fast Fourier transform, and the spectral edge frequency 95% (SEF95) was derived. SEF95 is defined as the frequency below which 95% of the signal power resides.

#### 6. Suppression ratio

The ratio of total suppression waves on BS pattern to 64 s was measured on EEG waveforms by manual inspection (Fig. 5g). A suppression wave was defined as a flattish wave lasting 0.35 s or more (*31*) with a spike wave appearing immediately after.

#### 7. Statistical analysis

We performed agglomerative hierarchical clustering for N**τ** to identify vigilance states. First, we computed the Euclidean distances as the dissimilarity values by the "dist" function and then fed these values into the "hclust" function with Ward’s minimum variance method "ward.D2". Next, we cut the dendrogram into 2 groups (awake and sleep) with the "cutree" function. The sleep group in $N\tau$ was further classified into 3 groups (NREM, REM, and light sleep) by the "hclust" function with the "centroid" or "average" method followed by the "cutree" function.

The mean value of the index in each state was used as the representative value for the individual. Since analysis data with 4 different sampling frequencies were collected at M2 and S1 LFPs, 8 representative values were set for each state for statistical examination. Wilcoxon signed-rank test was used to compare indices between M2 and S1. Holm-corrected exact Wilcoxon signed-rank test was used to compare new indices among vigilance states. After visual confirmation of a linear relationship between the indices and the SEV, Pearson correlation coefficient was calculated to quantify the strength and direction of the relationship.

The “paradoxical increase” is defined as both the mean value and two or three of the three thresholds being greater than those values at lower SEVs.

Values are expressed as the mean ± standard deviation or median. Statistical significance was defined as a p value $< 0.05$ for all the data. All the statistical analyses were performed using R software (version 3.6.2) with custom-made scripts. The graphical outputs were generated using gnuplot (version 5.1, http://www.gnuplot.info/).

**Acknowledgments**

I would like to express my deepest appreciation to Haruna Taichi at Tokyo Woman’s Christian University for significant contributions to writing the C script and valuable comments for analysis. I would also like to express my deepest appreciation to Masanori Murayama and his colleagues at RIKEN Center for Brain Science for their significant contributions to the collection of mouse data and the conception of this work. I also gratefully acknowledge Satoshi Hagihira at Kansai Medical University and Kazuto Yamashita at Rakuno Gakuen University, who supported the anesthesia study in dogs.

**Competing Interests statement**

The author declares no competing interests.

## Figures

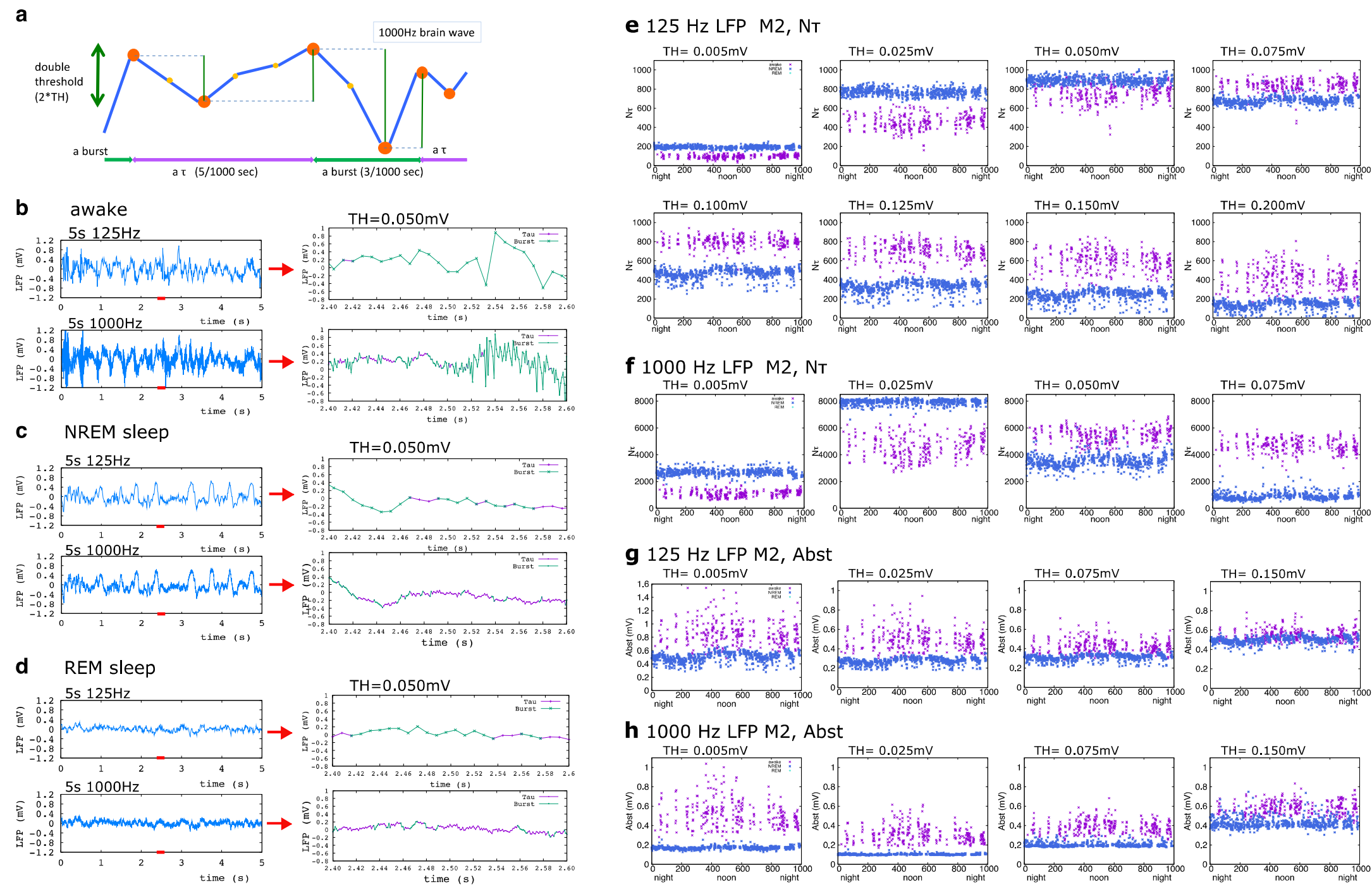

**Fig. 1. Concept of τ and burst and typical examples of changes in new indices with changes in the threshold on LFP at point No. 1.**

(**a**) Schematic diagram of **τ** and **burst** events. A **τ** is defined as a consecutive part of brain waves in which the potential difference between adjacent peaks is within a double threshold. A **burst** is defined as a consecutive part of brain waves in which it is above a double threshold. (**b-d**) Representative 5 s LFP at 125 Hz and 1000 Hz during awake, NREM, and REM in No. 1 (left). Examples of **τ** (purple) and **burst** (green) with 0.05 mV threshold are shown in the enlarged view (right). (**e-h**) Distribution of the number of **τ** (N**τ**) for trial number at various thresholds on 125 Hz LFP (e) and 1000 Hz LFP (f) at the secondary motor cortex (M2). Distribution of the amplitude of **burst** (Abst) for trial number at various thresholds on 125 Hz LFP (g) and 1000 Hz LFP (h) at M2. With increasing threshold, the changes in the indices in sleep state (*; n = 684) preceded those in awake state (×; n = 307) (No. 1), whose states were determined by the existing method.

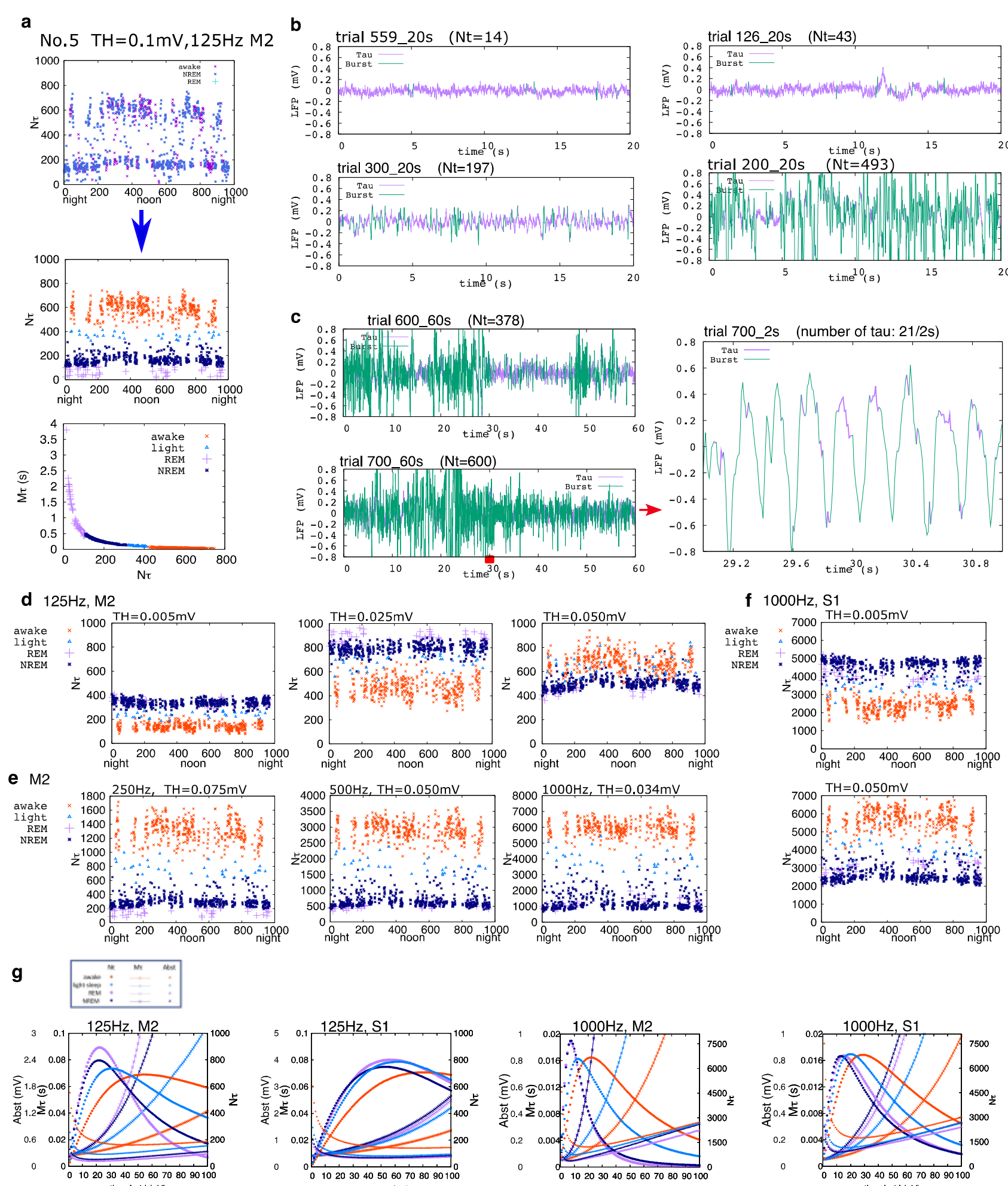

**Fig. 2. Results of hierarchical clustering of $N\tau$ for four vigilance-state classifications using 125 Hz LFP at M2 with 0.1 mV threshold at No. 5.**

**a**, Scatterplots of $N\tau$ on 125 Hz LFP at M2 with 0.1 mV threshold for trial number showing classification results using the existing method (top; ×: awake (n=323), *: NREM (n=646), +: REM (n=0)) and clustering-$N\tau$ method (center; ×: awake (n=414), △: light sleep (n=34), **+:** REM (n=50), and *: NREM (n=471)). Scatterplots of $M\tau$ for $N\tau$ (bottom)**. b**, **c**, Example of **τ** (purple) and **burst** (green) waveforms with 0.1 mV threshold on 20 s LFP (b), 60 s LFP (c_left), and 2 s LFP (c_right). **d**, Scatterplots of $N\tau$ for trial numbers using 125 Hz LFP at M2 with a threshold of 0.005 mV (left), 0.025 mV (center), or 0.050 mV (right). **e**, Scatterplots of $N\tau$ for trial number using 250 Hz LFP at M2 with 0.075 mV threshold (left), 500 Hz LFP with 0.050 mV threshold (center), and 1000 Hz LFP with 0.034 mV threshold (right). **f**, Scatterplots of $N\tau$ for trial number using 1000 Hz LFP at S1 with 0.005 mV (top) or with 0.050 mV (bottom) threshold. **g**, Relationships between threshold and mean $N\tau$ (●), $M\tau$ (×), or Abst (▲) in awake (orange), light sleep (light blue), REM (purple), and NREM (navy) states on 125 Hz and 1000 Hz LFP at M2 (left) and the primary somatosensory cortex (S1) (right).

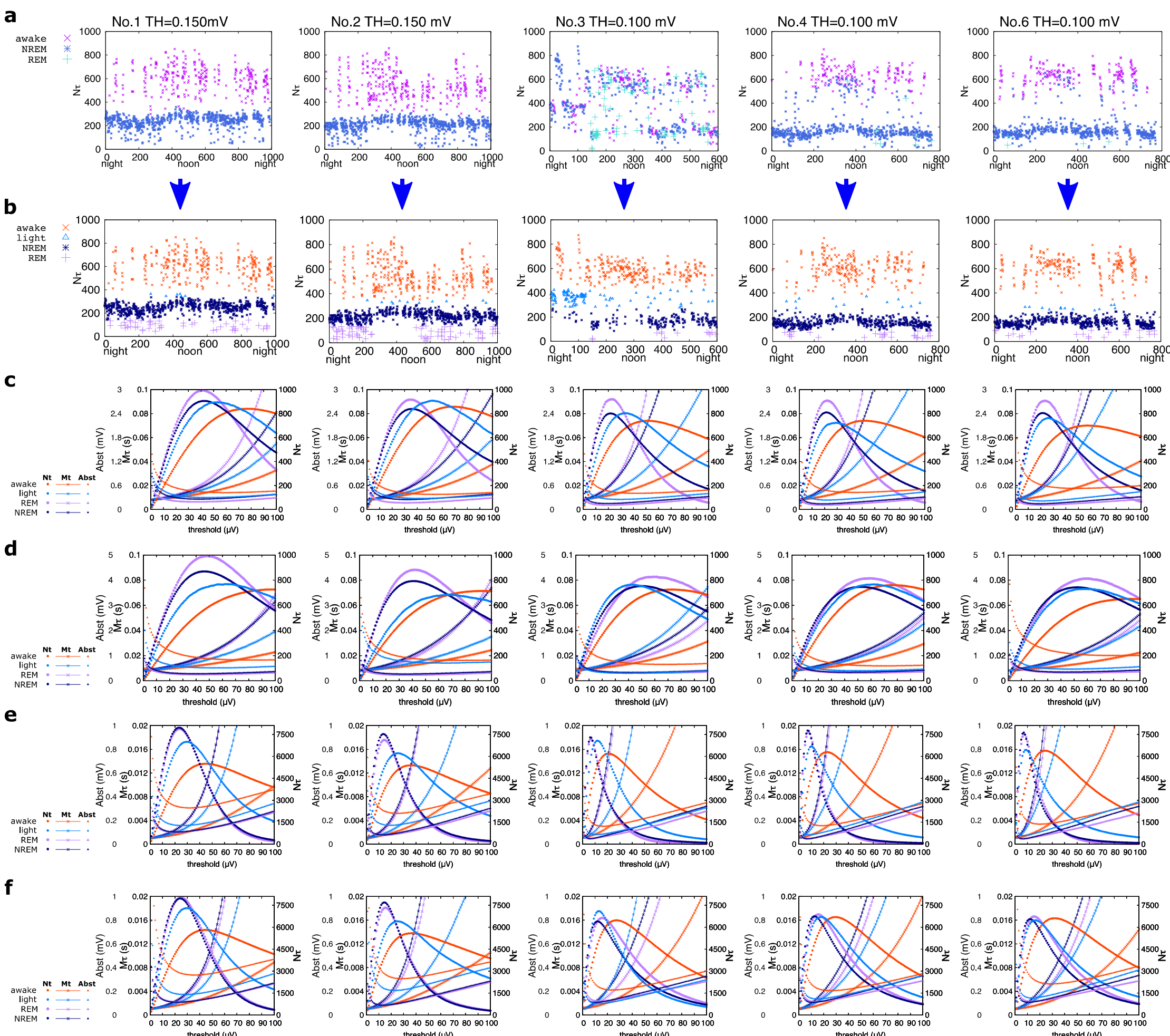

**Fig. 3. Results of hierarchical clustering to Nτ for four-state classification in Nos. 1-6, except for No. 5.**

**a**, **b**, Scatterplots of Nτ on 125 Hz LFP at M2 for trial number showing classification results using existing method (top; ×: awake, *: NREM, +: REM) (**a**) and clustering-Nτ method (bottomz; ×: awake, △: light sleep, **+:** REM, and ***:** NREM) (**b**). From left to right: No. 1 (threshold=0.15 mV), No. 2 (0.15 mV), No. 3 (0.1 mV), No. 4 (0.1 mV), and No. 6 (0.1 mV). **c**-**f**, Relationships between threshold value and mean Nτ (●), mean Mτ (×), or mean Abst (▲) in awake (orange), light sleep (light blue), REM (purple), and NREM (navy) states on 125 Hz LFP at M2 (**c**), on 125 Hz LFP at S1 (**d**), on 1000 Hz LFP at M2 (**e**), and on 1000 Hz LFP at S1 (**f**).

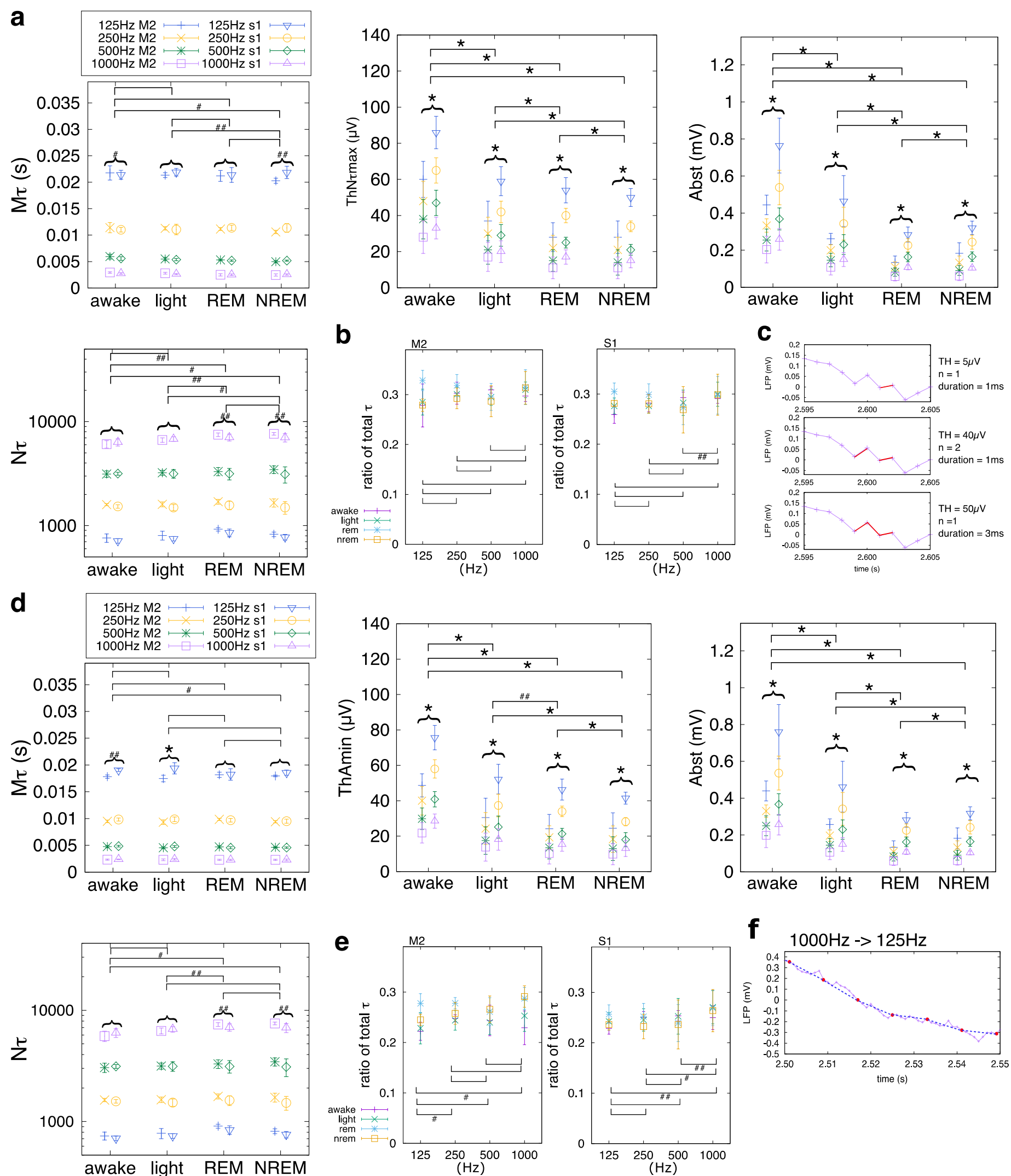

**Fig. 4. τ and burst at Mτ of the 2-3 sample intervals where Nτ is maximal and Abst is minimal in the 4 vigilance states**

**a**, Mean and SD of indices for maximizing mean Nτ (ThNtmax) in awake, light sleep, REM, and NREM states. Mτ for states: The mean Nτ is maximal when the mean Mτ is a 2.5-2.9 sample intervals, ThNτmax for states, Abst for states, and Nτ for states. **b**, Ratio of total τ at ThNτmax for sampling frequencies at M2 and S1. **c**, Specific example of Mτ when the number of τ changes. **d**, Mean and SD of indices for minimizing the mean Abst (ThAmin) in awake, light sleep, REM, and NREM states. Mτ for states: The mean Abst is the minimum when the mean Mτ is 2.2-2.4 sample intervals, ThAmin for states, Abst for states, and Nτ for states. **e**, Ratio of total τ at ThAmin for sampling frequencies at M2 and S1. **f**, Slow waveform at 1000 Hz (purple) and 125 Hz (blue). As the sampling frequency decreases, the microwaves on the slow wave disappear. Statistical significance among the vigilance state (n=48) and among M2 and S1 (n=24) was assessed by exact Wilcoxon signed-rank test with Holm correction; *$p<10^{-5}$, ##$p<0.001$, #$p<0.05$.

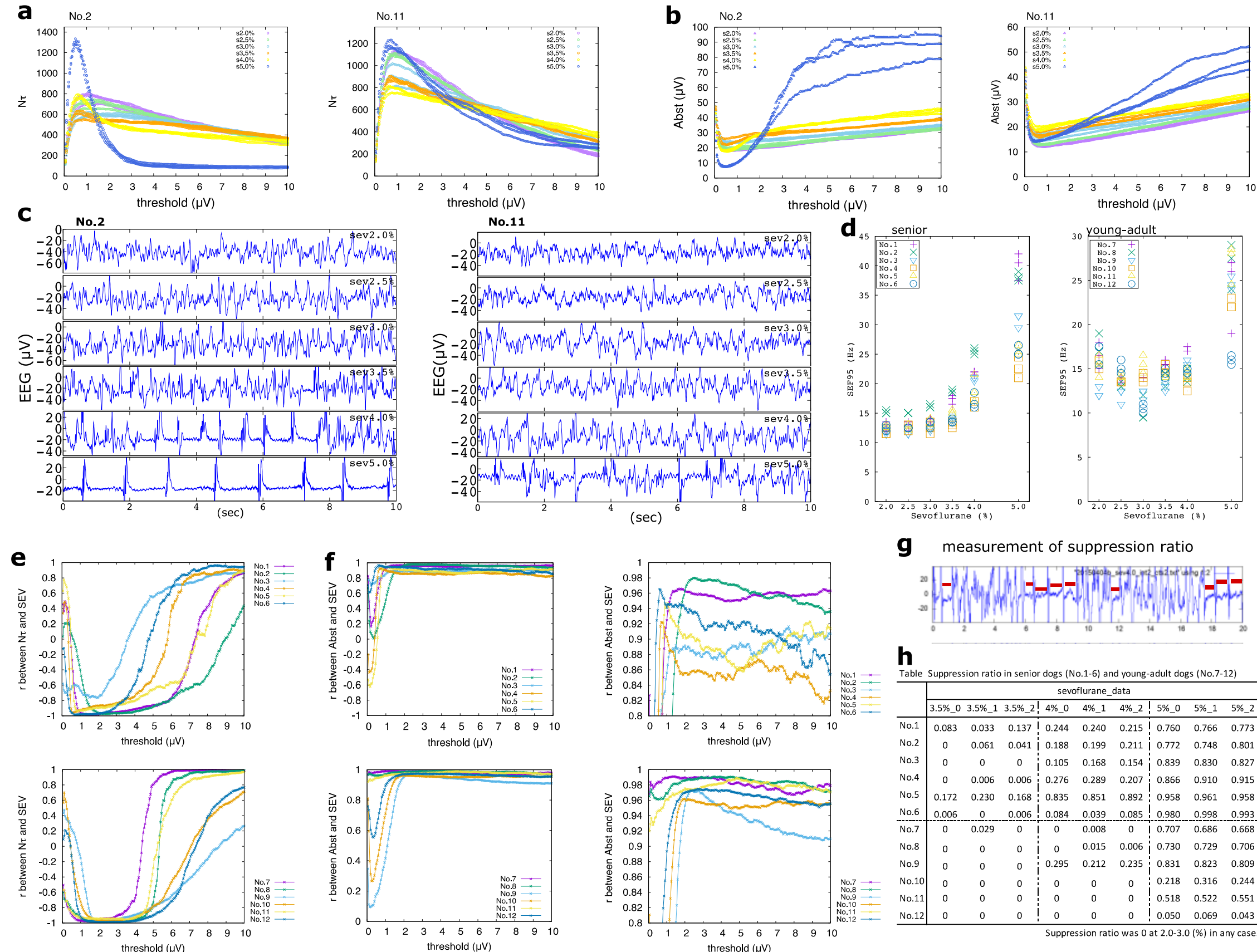

Table Suppression ratio in senior dogs (No.1-6) and young-adult dogs (No.7-12)

| | sevoflurane_data | | | | | | | | |
|---|---|---|---|---|---|---|---|---|---|
| | 3.5%_0 | 3.5%_1 | 3.5%_2 | 4%_0 | 4%_1 | 4%_2 | 5%_0 | 5%_1 | 5%_2 |
| No.1 | 0.083 | 0.033 | 0.137 | 0.244 | 0.240 | 0.215 | 0.760 | 0.766 | 0.773 |
| No.2 | 0 | 0.061 | 0.041 | 0.188 | 0.199 | 0.211 | 0.772 | 0.748 | 0.801 |
| No.3 | 0 | 0 | 0 | 0.105 | 0.168 | 0.154 | 0.839 | 0.830 | 0.827 |
| No.4 | 0 | 0.006 | 0.006 | 0.276 | 0.289 | 0.207 | 0.866 | 0.910 | 0.915 |
| No.5 | 0.172 | 0.230 | 0.168 | 0.835 | 0.851 | 0.892 | 0.958 | 0.961 | 0.958 |
| No.6 | 0.006 | 0 | 0.006 | 0.084 | 0.039 | 0.085 | 0.980 | 0.998 | 0.993 |
| No.7 | 0 | 0.029 | 0 | 0 | 0.008 | 0 | 0.707 | 0.686 | 0.668 |
| No.8 | 0 | 0 | 0 | 0 | 0.015 | 0.006 | 0.730 | 0.729 | 0.706 |
| No.9 | 0 | 0 | 0 | 0.295 | 0.212 | 0.235 | 0.831 | 0.823 | 0.809 |
| No.10 | 0 | 0 | 0 | 0 | 0 | 0 | 0.218 | 0.316 | 0.244 |
| No.11 | 0 | 0 | 0 | 0 | 0 | 0 | 0.518 | 0.522 | 0.551 |
| No.12 | 0 | 0 | 0 | 0 | 0 | 0 | 0.050 | 0.069 | 0.043 |

Suppression ratio was 0 at 2.0-3.0 (%) in any case.

**Fig. 5. Nτ and Abst at 0.05 μV-10 μV threshold on EEG in dogs under various sevoflurane doses.**

**a-c**, Typical results of N**τ** (**a**) and Abst (**b**) for threshold and EEG waveforms (**c**) in No. 2 of senior (left) and No. 11 of young adult (right) anesthetized with sevoflurane (SEV) at 2.0-5.0 (%). As SEV increases, the peak of the maximum N**τ** decreases at non burst suppression (BS) level and increases at BS level (**a**). The peak of the minimum Abst increases at non-BS level and decreases at BS level (**b**). EEG waveforms; BS level is 3.5% or more in No. 2 and 5% in No. 11 (**c**). **d**, Results of spectral edge frequency 95% (SEF95) in seniors (left) and in young adults (right). **e**, **f**, Results of correlation coefficients between N**τ** (e) or Abst (f) and SEV with suppression ratio <0.32 for each threshold in seniors (top) and in young adults (bottom). The right images are enlarged views (f). **g**, Example of the measurement of suppression ratio. **h**, Results of suppression ratio with SEVs ≥3.5%.

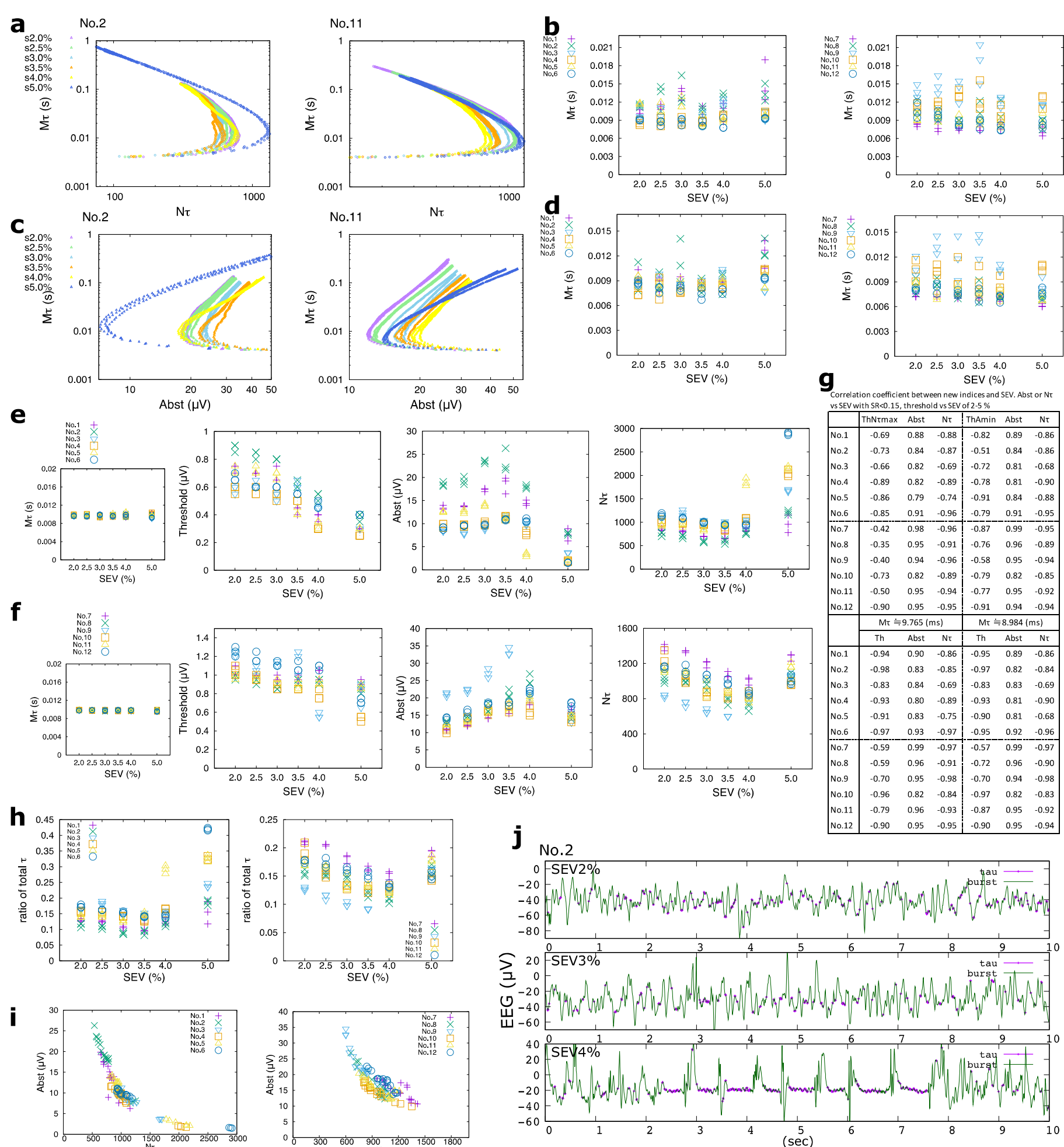

Correlation coefficient between new indices and SEV. Abst or Nτ vs SEV with SR<0.15, threshold vs SEV of 2-5 %

| | ThNτmax | Abst | Nτ | ThAmin | Abst | Nτ |
|---|---|---|---|---|---|---|
| No.1 | -0.69 | 0.88 | -0.88 | -0.82 | 0.89 | -0.86 |
| No.2 | -0.73 | 0.84 | -0.87 | -0.51 | 0.84 | -0.86 |
| No.3 | -0.66 | 0.82 | -0.69 | -0.72 | 0.81 | -0.68 |
| No.4 | -0.89 | 0.82 | -0.89 | -0.78 | 0.81 | -0.90 |
| No.5 | -0.86 | 0.79 | -0.74 | -0.91 | 0.84 | -0.88 |
| No.6 | -0.85 | 0.91 | -0.96 | -0.79 | 0.91 | -0.95 |
| No.7 | -0.42 | 0.98 | -0.96 | -0.87 | 0.99 | -0.95 |
| No.8 | -0.35 | 0.95 | -0.91 | -0.76 | 0.96 | -0.89 |
| No.9 | -0.40 | 0.94 | -0.96 | -0.58 | 0.95 | -0.94 |
| No.10 | -0.73 | 0.82 | -0.89 | -0.79 | 0.82 | -0.85 |
| No.11 | -0.50 | 0.95 | -0.94 | -0.77 | 0.95 | -0.92 |
| No.12 | -0.90 | 0.95 | -0.95 | -0.91 | 0.94 | -0.94 |

| | Mτ ≒9.765 (ms) | | | Mτ ≒8.984 (ms) | | |
|---|---|---|---|---|---|---|
| | Th | Abst | Nτ | Th | Abst | Nτ |
| No.1 | -0.94 | 0.90 | -0.86 | -0.95 | 0.89 | -0.86 |
| No.2 | -0.98 | 0.83 | -0.85 | -0.97 | 0.82 | -0.84 |
| No.3 | -0.83 | 0.84 | -0.69 | -0.83 | 0.83 | -0.69 |
| No.4 | -0.93 | 0.80 | -0.89 | -0.93 | 0.81 | -0.90 |
| No.5 | -0.91 | 0.83 | -0.75 | -0.90 | 0.81 | -0.68 |
| No.6 | -0.97 | 0.93 | -0.97 | -0.95 | 0.92 | -0.96 |
| No.7 | -0.59 | 0.99 | -0.97 | -0.57 | 0.99 | -0.97 |
| No.8 | -0.59 | 0.96 | -0.91 | -0.72 | 0.96 | -0.90 |
| No.9 | -0.70 | 0.95 | -0.98 | -0.70 | 0.94 | -0.98 |
| No.10 | -0.96 | 0.82 | -0.84 | -0.97 | 0.82 | -0.83 |
| No.11 | -0.79 | 0.96 | -0.93 | -0.87 | 0.95 | -0.92 |
| No.12 | -0.90 | 0.95 | -0.95 | -0.90 | 0.95 | -0.94 |

**Fig. 6. Threshold, Abst, and Nτ when Mτ was 2-3 sample intervals under 2%-5% SEV.** **a**, Typical results of Mτ for Nτ in No. 2 (left) and No. 11 (right) dogs anesthetized with SEV at 2.0 (purple), 2.5 (green), 3.0 (light blue), 3.5 (orange), 4.0 (yellow), and 5.0 (blue) (%) on a double-logarithmic graph. Nτ reached a maximum at all SEVs where Mτ was close to 0.01 s. **b**, Mτ for maximizing Nτ at SEV 2.0-5.0 (%) in seniors (Nos. 1-6) (left) and in young adults (Nos. 7-12) (right). **c**, Typical results of Abst for Nτ using No. 2 (left) and No. 11 (right) on a double-logarithmic graph. **d**, Mτ for minimizing Abst at each SEV in seniors (left) and in young adults (right). **e**, **f**, Mτ, threshold, Abst, and Nτ at each SEV, where Mτ was closest to 9.765 ms (left) in seniors (**e**) and in young adults (**f**). **g**, Table of correlation coefficients between SEV of 2-5% and the threshold and between SEV with SR<0.15 and Nτ or Abst, where Nτ was the maximum (ThNτmax), Abst was the minimum (ThAmin), and Mτ was closest to 9.765 ms and 8.984 ms. **h**, **i**, Ratio of total τ at each SEV (h) and Abst for Nτ (i), where Mτ was closest to 9.765 ms in seniors (left) and young adults (right). **j**, Example of τ and **burst** waveform where Mτ was closest to 9.765 ms in No. 2.

## Supplementary figures

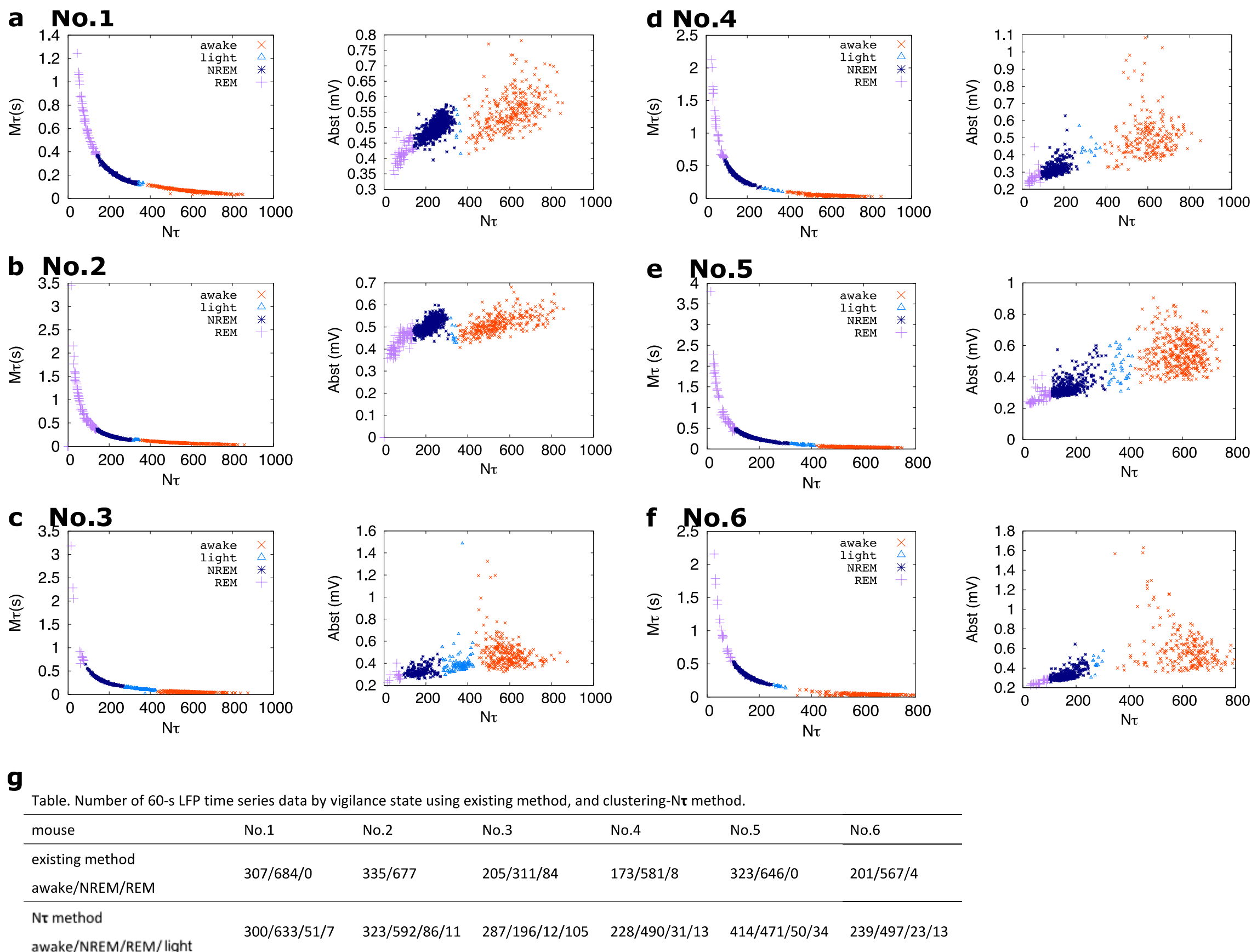

**g**

Table. Number of 60-s LFP time series data by vigilance state using existing method, and clustering-Nτ method.

| mouse | No.1 | No.2 | No.3 | No.4 | No.5 | No.6 |
|---|---|---|---|---|---|---|
| existing method<br>awake/NREM/REM | 307/684/0 | 335/677 | 205/311/84 | 173/581/8 | 323/646/0 | 201/567/4 |
| Nτ method<br>awake/NREM/REM/ light | 300/633/51/7 | 323/592/86/11 | 287/196/12/105 | 228/490/31/13 | 414/471/50/34 | 239/497/23/13 |

**h**

Mean ±SD of new parameters for awake and 3 stages of sleep using LFP sampled at 125hz at M2 with threshold of 0.1 mV or 0.15 mV in  6 mice.

| | awake | light | rem | nrem |
|---|---|---|---|---|
| Nt | 588±91 | 349±40 | 75±33 | 192±40 |
| Mt (ms) | 54±21 | 117±25 | 880±543 | 270±86 |
| Tt (%) | 51±14 | 67±8 | 88±12 | 80±5 |
| Abst (mcV) | 536±123 | 425±108 | 351±97 | 400±98 |

Nt = nuber of tau, Mt = mean duration of tau
Tt = ratio of total tau, Abst = amplitude of burst

**Fig. S1**

**a-f**, Scatterplots of Mτ for Nτ (left) and Abst for Nτ (right) on 125 Hz LFP at M2 with a threshold of 0.15 mV (**a**, **b**: Nos. 1, 2) or 0.10 mV (**c-f**: Nos. 3-6). The state was determined using the clustering-Nτ method. **g**, Results of the number of LFP data by vigilance state identification for each subject. **h**, Mean ± SD of Nτ, Mτ, the ratio of total τ (Tt), and Abst at 125 Hz at M2 with a 0.15 mV or 0.1 mV threshold in 6 mice.

Table Mean ± sd of SEF95 of various frequencies at M2 and S1 in each mice under awake and NREM, REM, and light sleeps .

| id | M2 | | | | S1 | | | |
|---|---|---|---|---|---|---|---|---|
| | awake | NREM | REM | light | awake | NREM | REM | light |
| | 125 Hz | | | | 125 Hz | | | |
| No.1 | 55±3 | 21±7 | 39±5 | 39±21 | 24±12 | 24±6 | 33±9 | 22±9 |
| No.2 | 53±4 | 17±6 | 33±11 | 49±11 | 25±14 | 23±5 | 29±11 | 24±15 |
| No.3 | 28±14 | 21±7 | 26±11 | 21±10 | 26±12 | 27±5 | 34±5 | 26±7 |
| No.4 | 27±13 | 23±6 | 37±8 | 24±11 | 29±11 | 27±4 | 33±7 | 29±8 |
| No.5 | 23±11 | 23±6 | 34±10 | 25±10 | 25±11 | 27±5 | 31±7 | 27±10 |
| No.6 | 23±11 | 22±5 | 35±8 | 24±7 | 21±8 | 25±4 | 33±6 | 22±9 |
| | 250 Hz | | | | 250 Hz | | | |
| No.1 | 111±7 | 25±16 | 64±13 | 69±49 | 39±30 | 28±10 | 47±15 | 26±17 |
| No.2 | 106±9 | 20±12 | 46±20 | 94±26 | 40±33 | 25±8 | 37±17 | 37±34 |
| No.3 | 44±32 | 22±10 | 31±18 | 27±22 | 33±25 | 29±8 | 38±8 | 30±14 |
| No.4 | 42±29 | 26±9 | 50±16 | 26±13 | 37±20 | 29±5 | 37±9 | 33±11 |
| No.5 | 30±21 | 25±10 | 44±18 | 31±18 | 30±19 | 28±6 | 35±10 | 32±15 |
| No.6 | 32±24 | 23±6 | 47±17 | 27±11 | 23±11 | 26±5 | 38±9 | 23±10 |
| | 500 Hz | | | | 500 Hz | | | |
| No.1 | 226±10 | 34±36 | 103±32 | 133±107 | 77±74 | 34±20 | 60±23 | 39±48 |
| No.2 | 218±13 | 23±27 | 61±40 | 193±60 | 78±78 | 27±17 | 43±25 | 66±82 |
| No.3 | 81±74 | 24±15 | 34±21 | 37±48 | 50±55 | 30±11 | 40±9 | 36±30 |
| No.4 | 73±68 | 28±16 | 57±22 | 28±14 | 52±45 | 29±6 | 38±10 | 34±13 |
| No.5 | 45±48 | 28±17 | 48±21 | 44±43 | 38±35 | 29±8 | 36±11 | 36±23 |
| No.6 | 51±56 | 24±8 | 54±30 | 35±21 | 26±19 | 27±5 | 39±10 | 24±12 |
| | 1000 Hz | | | | 1000 Hz | | | |
| No.1 | 421±22 | 51±77 | 176±76 | 245±212 | 159±136 | 53±50 | 107±45 | 60±92 |
| No.2 | 414±26 | 30±56 | 90±90 | 369±120 | 151±146 | 34±38 | 56±50 | 123±162 |
| No.3 | 164±152 | 31±31 | 48±22 | 59±95 | 87±115 | 32±17 | 42±11 | 46±58 |
| No.4 | 147±143 | 35±33 | 75±43 | 38±19 | 86±96 | 31±7 | 40±11 | 43±18 |
| No.5 | 91±115 | 33±32 | 54±24 | 72±90 | 59±77 | 30±7 | 38±12 | 41±28 |
| No.6 | 103±123 | 28±16 | 70±54 | 64±67 | 34±40 | 28±5 | 41±10 | 26±12 |

**Fig. S2**

Table of the mean ± SD of spectral edge frequency 95% (SEF95) on 125, 250, 500, and 1000 (Hz) LFPs at M2 and S1 in awake, NREM, REM, and light sleep states of Nos. 1-6 mice.

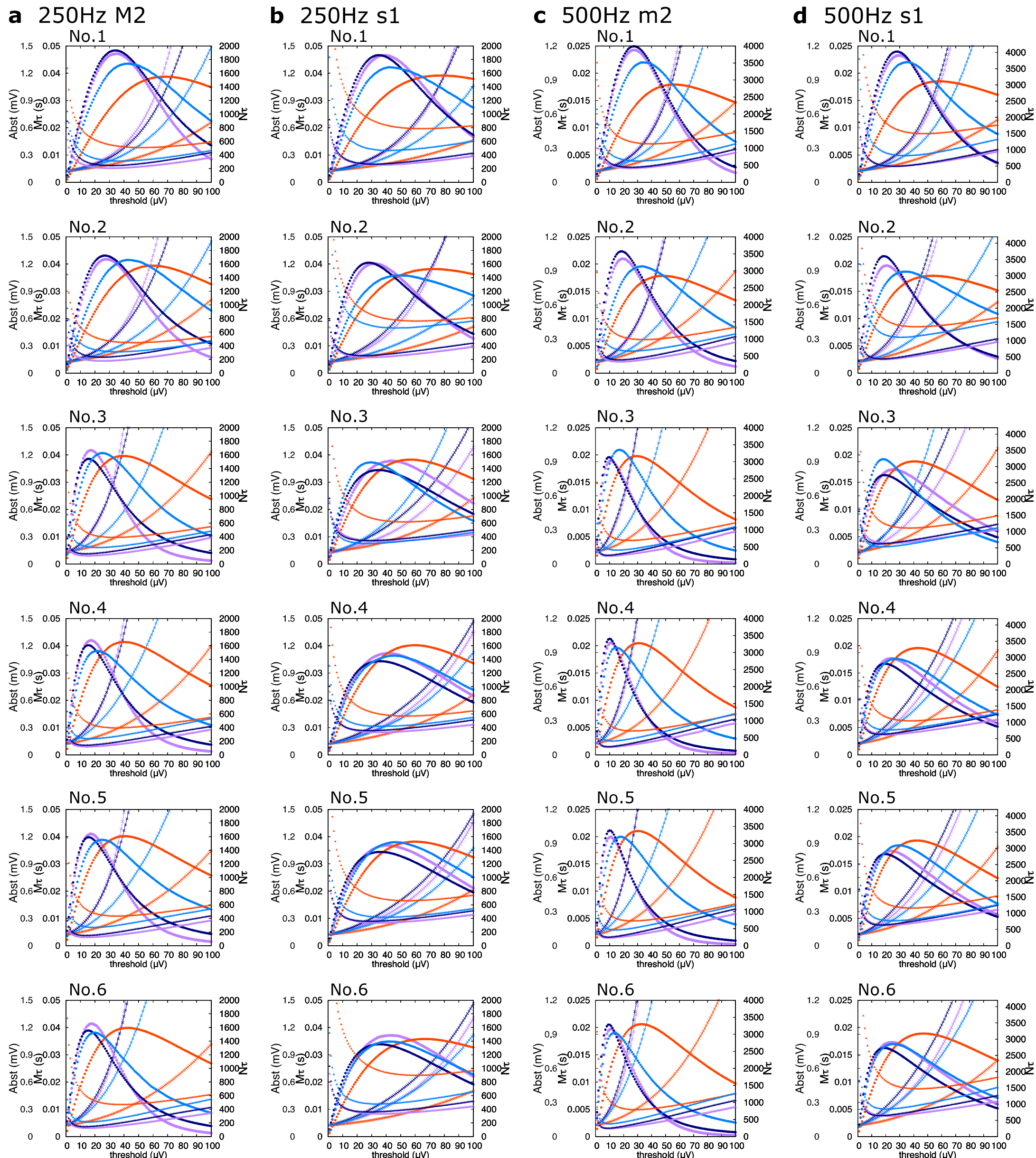

**Fig. S3**

Relationship between threshold value and mean Nτ (circle), mean Mτ (cross), or mean Abst (triangle) in awake (orange), light sleep (light blue), REM (purple), and NREM (navy) states on 250 Hz LFP at M2 (**a**), on 250 Hz LFP at S1 (**b**), on 500 Hz LFP at M2 (**c**), and on 500 Hz LFP at S1 (**d**). From top to bottom: No. 1, No. 2, No. 3, No. 4, No. 5, No. 6.

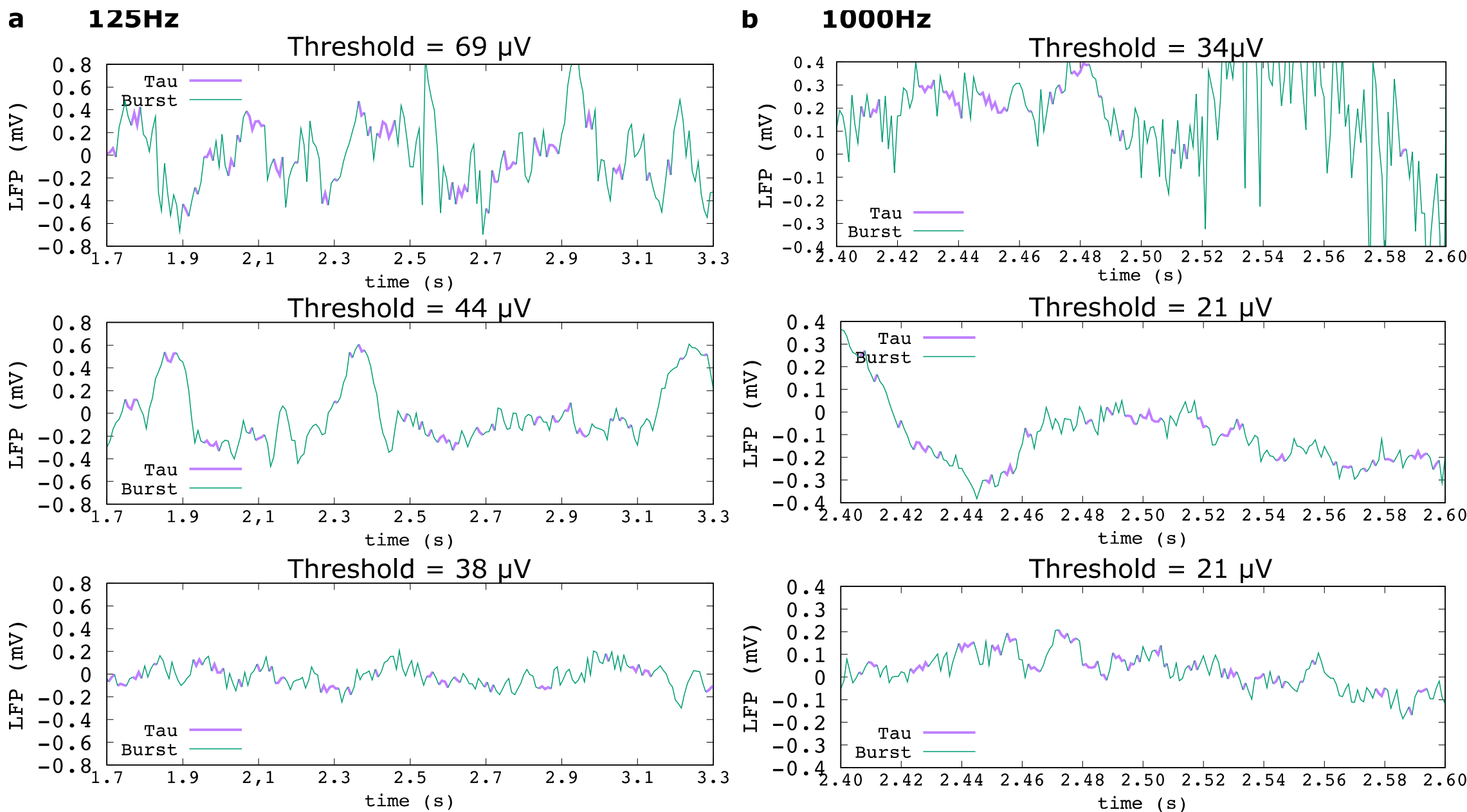

**Fig. S4**

Examples of **τ** (purple) and **burst** (green) waveforms showing self-similarity at a threshold for M**τ** of 2.5 sample intervals (same as Fig. 1b-d). **a**, Waveform of 1.6 sec LFP at 125 Hz (200 samples) with a vertical range from -0.8 mV to 0.8 mV in awake (top), NREM (center), and REM (bottom) states. **b**, Waveform of 0.2 sec LFP at 1000 Hz (200 samples) with a vertical range from -0.4 mV to 0.4 mV in awake (top), NREM (center), and REM (bottom) states.

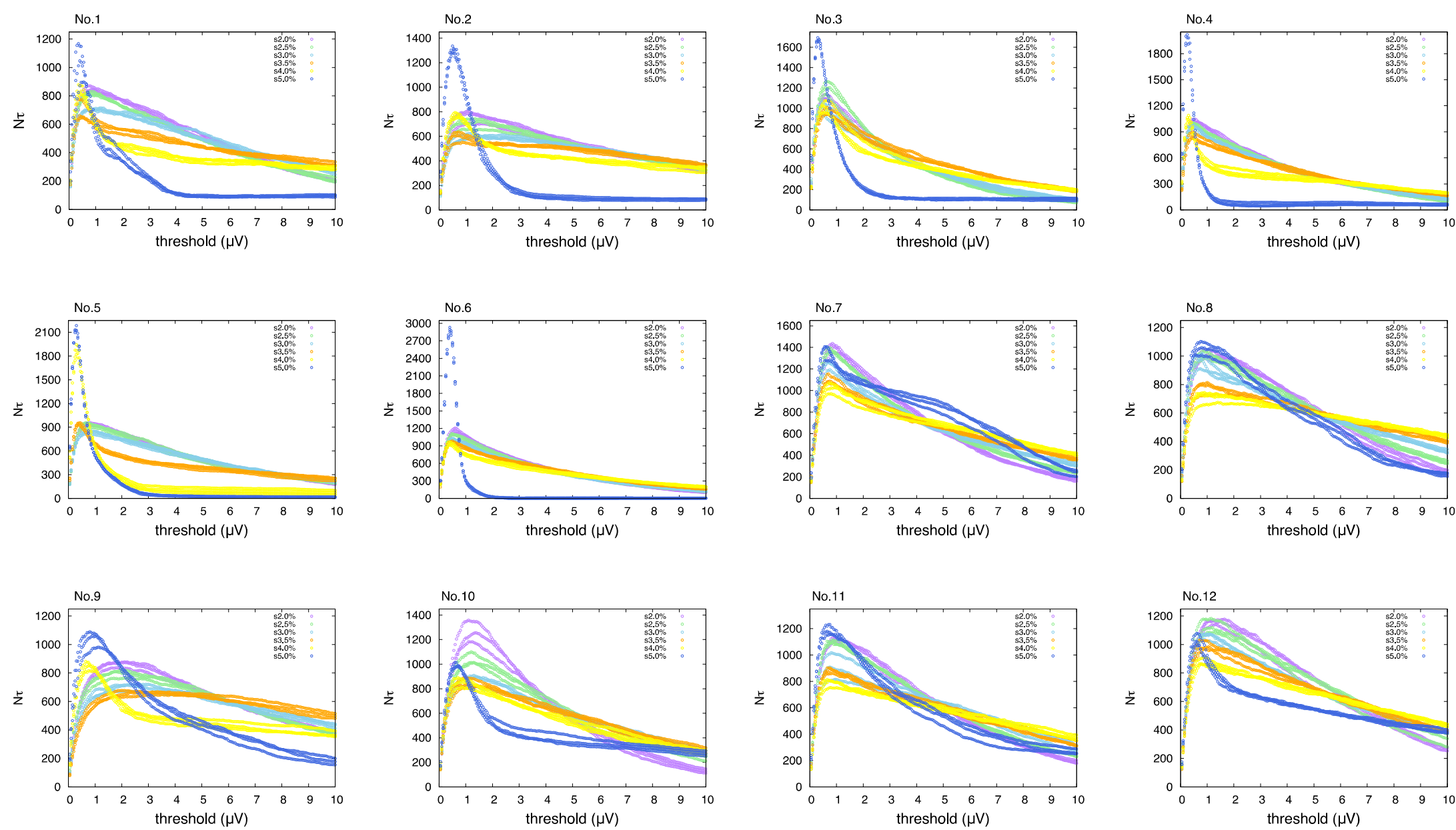

**Fig. S5**

**Nτ** for the threshold every 0.05 mV from 0.05 mV to 10 mV on EEG in Nos. 1-12 anesthetized with sevoflurane 2.0, 2.5, 3.0, 3.5, 4.0, and 5.0 (%).

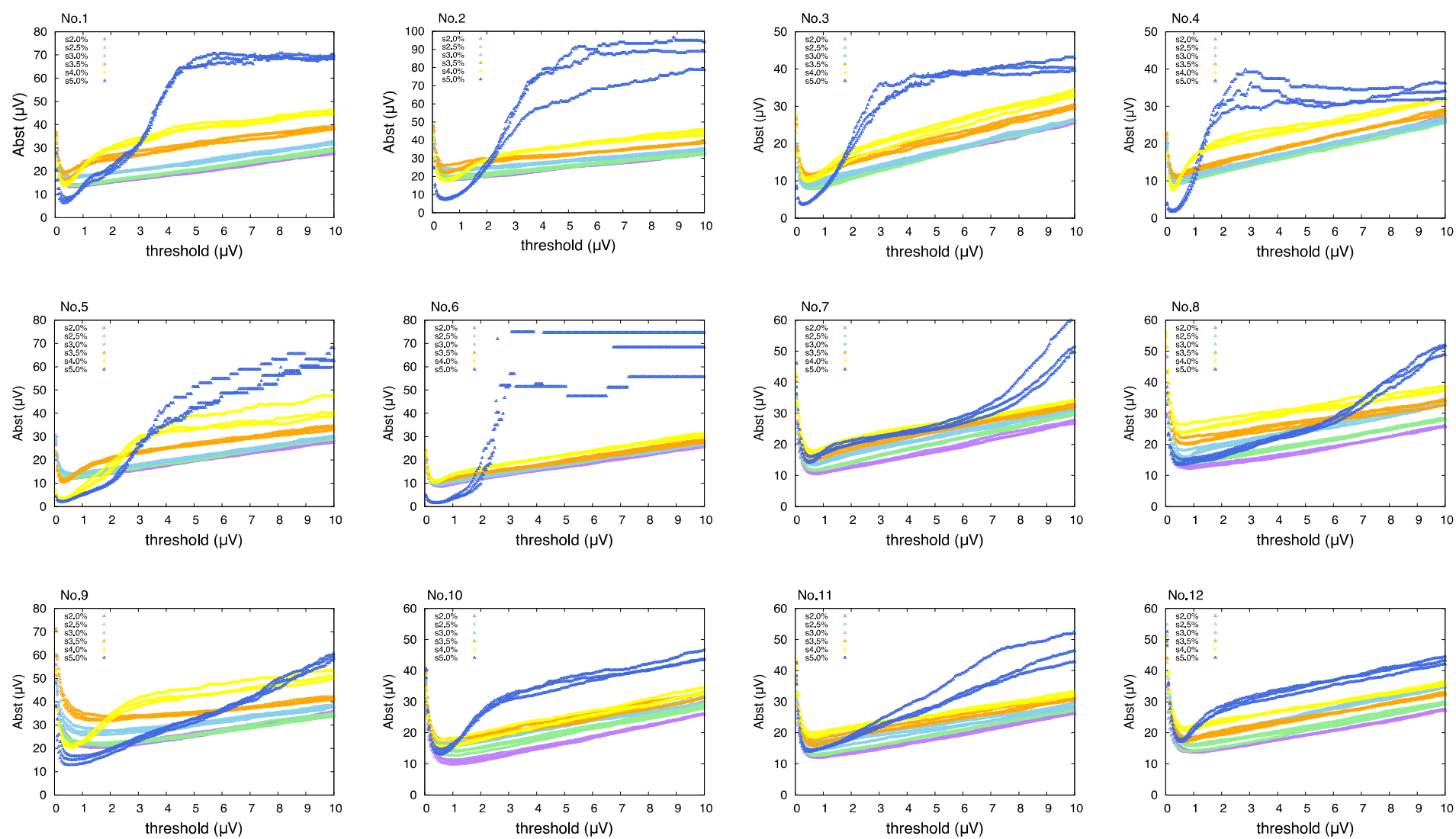

**Fig. S6**

Abst for the threshold at every 0.05 mV increase from 0.05 mV to 10 mV on EEG in Nos. 1-12 anesthetized with sevoflurane 2.0, 2.5, 3.0, 3.5, 4.0, and 5.0 (%).